\documentclass[12pt]{article}
\usepackage{amsmath,amssymb,color,appendix}
\usepackage{slashed}
\usepackage{bbm}
\usepackage{color}
\usepackage{chngpage}
\usepackage{graphicx}
\usepackage{dsfont}
\usepackage{upgreek}
\usepackage{color}
\usepackage{soul}
\usepackage{cancel}
\usepackage[normalem]{ulem}
\usepackage{float}
\usepackage{url}
\definecolor{gre}{rgb}{0,0.4,0.3}

\newcommand{\be}{\begin{equation}}
\newcommand{\eqb}{\begin{eqnarray}}
\newcommand{\eqf}{\end{eqnarray}}
\newcommand{\bb}{\begin{equation}}
\newcommand{\ee}{\end{equation}}
\newcommand{\beq}{\begin{equation}}
\newcommand{\eeq}{\end{equation}}
\newcommand{\bea}{\begin{eqnarray}}
\newcommand{\eea}{\end{eqnarray}}

\newcommand{\sd}{\!\!\not\!\partial}
\newcommand{\sA}{\!\not\!\! A}
\definecolor{gre}{rgb}{0,0.4,0.3}
\usepackage{chngpage}

\usepackage{color}

\title{From $D=3$ to $D=2$ dimensions: a
note on topological order }
\author{C.~D.~Fosco$^a$ and F.~A.~Schaposnik$^b$
\\
~
\\
~
\\
{\normalsize $^a\!$\it Centro At\'omico Bariloche and Instituto
Balseiro,}\\
{\normalsize $\!$\it Comisi\'on Nacional de Energ\'\i a At\'omica, 8400
Bariloche, Argentina}\\
{\normalsize $\!$\it }\\
{\normalsize $^b\!$\it Departamento de F\'\i sica, Universidad
Nacional de La Plata}\\ {\normalsize\it Instituto de F\'\i sica La Plata-CONICET}\\
{\normalsize\it C.C. 67, 1900 La Plata,
Argentina}
 }
\begin{document}
\date{\today}
\maketitle
\begin{abstract}
We construct, by a procedure involving a dimensional reduction from a Chern-Simons theory with borders, an effective theory for $1+1$ dimensional superconductor. That system can be either in an ordinary phase or in a topological one, depending on the value of two phases, corresponding to 
complex order parameters.
Finally, we argue that the original theory and its dimensionally reduced one can be related to the effective action for a quantum Dirac field in a slab geometry, coupled to a gauge field.
\end{abstract}
\section{Introduction}
Topology and quantum field theory intertwine in many different situations.
This is partly a consequence of quantum fluctuations, which are strongly
dependent on the spatial geometry and boundary conditions; besides, they also probe the geometry of the space of configurations, leading to quantum corrections that are naturally influenced by the properties of that space.

That interplay manifests itself in many different fashions: a first aspect,
on which we focus in this note, arises in the study of quantum matter
models where order parameters characterizing quantum states are determined by topological quantities. Among the condensed matter systems that can be successfully described in terms of a so termed ``topological order'', one finds topological insulators and superconductors (see, for example,~\cite{caQZ} and references in there).
A second instance is that of topological stability, a subject that has played
an important role in studies regarding general features of quantum field
theory. Indeed, in the '70$\,$s,  this very same property was the focus of
intense research, when dealing with non-perturbative quantization around
non-trivial classical configurations.  Field theories allowing for
soliton-like classical solutions (namely: kinks, vortices and monopoles in
$d=1,\,2$ or $3$ spatial dimensions, or instantons in their $D=d+1$ Euclidean space-time counterparts) were at the basis of WKB-like path-integral quantization methods  (see~\cite{Coleman} and references therein).
Finally, in yet another kind of development, there have been elaborations
on the so-called topological quantum field theories, i.e., systems with the
distinctive characteristic of having correlation functions which depend
only on global features of the space on which they are defined~\cite{Bir}.

In this note, we deal mostly with the first aspect, namely, topological
order systems: we consider a $D=3$ Chern-Simons (CS) model and its
dimensional reduction to a $D=2$ ``axion model''. Working within the
path-integral framework, particularly in its  treatment of  chiral
anomalies, we wish to describe different features of topological
systems, discussed in various frameworks~\cite{HZ}-\cite{QiWi}.
 Our analysis relies heavily on that approach, since it allows one to describe models defined on manifolds with borders in a rather straightforward way. It also makes it possible, for example, to identify topological aspects  with the associated non-invariance of path-integral  measures under symmetry transformations which have quantum anomalies.

This paper is organized as follows: as  a first step in the construction of a $2$-dimensional model, we consider, in Sect.~\ref{sec:borders}, a pure Abelian CS theory in the presence of borders and of an external source.  We find, in that context, the properties that we then use in Sect.~\ref{sec:them} to derive a dimensionally reduced (to $D=2$) model in which one can explicitly confirm the realization of a topological order. The latter amounts to augmenting the number of degrees of freedom in the reduced theory; i.e., on the boundaries, while keeping the symmetries found in the original theory.

Other aspects of the relationship between topology and quantum field theory also manifest themselves here; indeed, the fact that the starting point is a topological field theory, as well as the existence of non-trivial, vortex-like solutions in the reduced theory. These solutions are also studied in Sect.~\ref{sec:them}.

In Sect.~\ref{sec:fer} we consider a possible mechanism, starting from a Dirac field in $2+1$ dimensions, which leads to an effective action similar to the one considered in this note.

In Sect.~\ref{sec:conc} we present our conclusions as well as a discussion of our results.

\section{Abelian Chern-Simons theory in the presence of two planar parallel
borders}\label{sec:borders}
We shall begin our study by considering an Abelian Chern-Simons   action,
defined on a 3-dimensional manifold $U$, which has a non-trivial boundary
${\mathcal M} \equiv \partial U$. As stated in the Introduction, we are interested here in a case where that boundary corresponds to two parallel planes: the two-dimensional region spanned by two parallel static straight lines during time evolution.

In our conventions, coordinates in $3$ dimensions are denoted by $x^a$ (letters from the beginning of the Roman alphabet run over the values $1$, $2$ and $3$). The manifold $U$ will be assumed to be $U =\{(x^1,x^2,x^3): 0 < x^3 < \ell \}$; thus, ${\mathcal M}$ consists of two parallel planes, which we denote by $L$ and $R$, and correspond to $x^3= \ell$ and $x^3=0$, respectively. An Abelian CS action, in the absence of borders, including the coupling to a conserved current $s^a$, can be written as follows:
\be
S(A,s) \;=\; \frac1{4\pi} \int d^3x \, \varepsilon^{abc} \, A_a \partial_b A_c \;-\;
\int d^3x \, s^a A_a \;,
\label{CSJ}
\ee
with $A_a$ denoting a $U(1)$ gauge field, and $a = 1, 2, 3$.

The classical field equations for this action may be put in the form
\begin{equation}\label{eq:eeqq}
{\mathcal J}^a(x) \;=\; s^a(x) \;,
\end{equation}
where we have introduced the `Chern-Simons current' ${\mathcal J}^a(x)$,
\begin{equation}
{\mathcal J}^a(x) \;=\; \frac{1}{2\pi} \, \varepsilon^{abc} \, \partial_b A_c(x) \;.
\end{equation}

The CS current is conserved, $\partial_a {\mathcal J}^a = 0$, and ${\mathcal J}_3$ is related to the $2$-dimensional  metric-independent factor $\varepsilon^{ij} F_{ij}d^2x$, the integrand of the topological Chern number $C_1$.
 Therefore,
when constructing a generating functional of CS current correlation functions in a system with borders, ${\mathcal Z}(s)$, it is natural to impose the vanishing of the normal component of that current. This satisfies the integral form of the Gauss law, derived from the conservation of the current, at the boundary of $U$.

One might argue that this is not the most general way to satisfy the vanishing of the current flux; note, however, that if all the points at the boundary are equivalent, that is the only consistent choice (compatible with a vanishing total charge emerging from the system).

Therefore, the generating functional ${\mathcal Z}(s)$ for the $2+1$ dimensional theory in the presence of the boundary ${\mathcal M}$ must include a functional $\delta$-function $\delta_{\mathcal M}(J_n)$ imposing the vanishing of the normal component of the current is
\begin{equation}\label{eq:defzs}
{\mathcal Z}(s) \;=\; \int {\mathcal D}A_a \;
\delta_{\mathcal M}(J_n) \; e^{i{\mathcal S}(A,s)} \;.
\end{equation}
Imposing that condition isolates the problem inside of $U$ from the exterior problem. Therefore, in the action for the gauge field we may include all of space-time, and not just the points inside $U$. That is the reason why we have used  in (\ref{eq:defzs}) the same action as in (\ref{CSJ}) (absence of boundaries).

 In our case, the border is composed of two planes, and the respective normals are $n_L = {\check e_3}$ and $n_R = -{\check e_3}$.  Now, following the approach of~\cite{Fosco:2018nio,Fosco:2018pcq}, the constraint on the normal component of the current on ${\mathcal M}$ can be conveniently introduced in terms of a functional Fourier representation, at the expense of using two
auxiliary scalar fields in $1+1$ dimensions, which we denote by $\xi_L({\mathbf x})$ and
$\xi_R({\mathbf x})$, ${\mathbf x}=(x^i)_{i=1}^2$
(letters from the middle of the alphabet run from $1$ to $2$)
\begin{align}\label{eq:defsm}
\delta_{\mathcal M}(J_n) &=\; \int {\mathcal D} \xi_L {\mathcal D} \xi_R \;
e^{i S_{\mathcal M}(\xi, {\mathcal J})} \;\;, \nonumber\\
S_{\mathcal M}(\xi, {\mathcal J}) &=\; \int d^2x \,
\left[\xi_L({\mathbf x}) \, {\mathcal J}^3({\mathbf x},\ell)
\;-\; \xi_R({\mathbf x}) \, {\mathcal J}^3({\mathbf x},0) \right] \;.
\end{align}

Using this representation for the functional $\delta$, we see that we have for the generating functional the expression
\begin{equation}\label{eq:defzss}
{\mathcal Z}(s) \;=\; \int {\mathcal D}\xi_L \,{\mathcal D}\xi_R \;
{\mathcal D}A \; e^{i{\mathcal S}(A,s+c)} \;,
\end{equation}
where we have taken advantage of the fact that the term involving the auxiliary fields may also be regarded as the result of adding an extra current, coupled to the gauge field
\begin{equation}\label{eq:defc}
c^j(x) \;=\; \delta(x^3) \;
\varepsilon^{jk} \,\partial_k\xi_R({\mathbf x})
 \,-\, \delta(x^3-\ell) \varepsilon^{jk} \,\partial_k\xi_L({\mathbf x}) \;\;,
 \;\;\; c^3(x) \;=\;0\;.
\end{equation}
Note that this new current $c^a$ is also conserved.

The partition function in (\ref{eq:defzss}) involves then an integration over $A$ and the two auxiliary fields, one on each face of the boundary. Let us first consider the integration over the gauge field. Since it is a Gaussian,
we know that the result of performing it may be put in the form:
\begin{equation}\label{eq:zss2}
{\mathcal Z}(s) \;=\; \int {\mathcal D}\xi_L \,{\mathcal D}\xi_R \;
\; e^{i{\mathcal S}(A,s+c)} \;,
\end{equation}
where $A$ is the classical solution for the gauge field in the presence of the current $s+c$, which is partly external and partly topological (we are ignoring a factor which is independent of the current).

Let us consider the example of an external source $s^a$ such that $s^j=0$,
and:
\begin{equation}\label{eq:sext}
s^3(x) \;=\; s^3({\mathbf x},x^3) \;,
\end{equation}
taking the same values on the two faces:
$s^3({\mathbf x},0) = s^3({\mathbf x},\ell) \equiv s^3({\mathbf x})$, so that the third component of the CS current is the same: $\frac{1}{2\pi}\varepsilon^{ij}\partial_j A_j({\mathbf x})$.

Since the gauge field in (\ref{eq:defzss}) is coupled to $s+c$, the classical equations of motion for the gauge field are:
\begin{equation}
{\mathcal J}^3(x) \;=\;\frac{1}{2\pi}\varepsilon^{ij}\partial_i A_j({\mathbf x},x^3)\;=\; s^3({\mathbf x},x^3) \;,\;\;
{\mathcal J}^i(x) \;=\; c^i (x) \;.
\end{equation}
Note that the current ${\mathcal J}^3$
\be
{\mathcal J}^3 = \frac1{4\pi} \varepsilon^{ij}F_{ij} \;, \hspace{2 cm}  i = 1,2 \;,
\label{11}
\ee
coincides with the axial anomaly for a $D=2$ massless Weyl fermion  since the factor  in Eq.~\eqref{11} is $1/2$ of the anomaly associated to a Dirac fermion.
Indeed, within the path-integral framework the chiral current anomaly is related to the Jacobian $J[\alpha]$ of a chiral transformation, with $\alpha$ denoting the chiral phase rotation~\cite{Fujikawa}. Now, in the case of Weyl fermions one  can prove that (see below)
\be
{\cal A}_{Weyl}(\alpha) = - \left.\frac{\delta \log J_L[\alpha]}{\delta\alpha}\right|_{\alpha =0} = \frac12 {\cal A}_D
\ee
where ${\cal A}_D$ is the   anomaly for Dirac fermions associated to a chiral transformation $\exp(i\gamma_5\alpha)$ which, in $D=2$, takes the form:
\be
{\cal A}_D = \frac1{2\pi}\int d^2x \varepsilon^{ij}F_{ij} \;.
\ee

Now, since a Majorana-Weyl fermion may be thought of as carrying half the number of degrees of freedom of a Weyl fermion, one may interpret this result as due the presence of two Majorana-Weyl fermions, with opposite helicity, at the edges.

In the $A_3 \equiv 0$ gauge the equation involving $c^i$ may be written more explicitly as follows:
\begin{equation}\label{eq:aux}
-\frac{1}{2\pi} \partial_3 A_i(x)  \;=\; \partial_i[\delta(x^3) \, \xi_R({\mathbf x})  - \delta(x^3-\ell) \, \xi_L({\mathbf x})] \;.
\end{equation}

Finally, recalling (\ref{eq:zss2}), we see that for this configuration:
\begin{equation}
{\mathcal S}[A,s+c] \,=\,-\frac{1}{2} \int d^3x \, s^a(x) A_a(x)
\,=\, -\frac{1}{2} \int d^3x \, c^j(x) A_j(x)
\end{equation}
since $A_3 =0$. Therefore,
\begin{equation}
{\mathcal S}[A,s+c] \,=\,\frac{1}{4\pi} \int d^2x \,
[\xi_L({\mathbf x}) -  \xi_R({\mathbf x})] \, \varepsilon^{ij}\partial_j A_j({\mathbf x}) \;.
\label{13}
\end{equation}
We have seen, therefore, that
\begin{equation}\label{eq:zss3}
{\mathcal Z}[A] \;=\; \int {\mathcal D}\xi_L \,{\mathcal D}\xi_R \;
\; e^{\frac{i}{4\pi} \int d^2x \,
[\xi_L({\mathbf x}) -  \xi_R({\mathbf x})] \, \varepsilon^{ij}\partial_j A_j({\mathbf x})}
\;.
\end{equation}

In the next Section we promote the auxiliary fields $\xi$ to dynamical ones, by equipping them with a non-trivial action.

\section{The model}\label{sec:them}
\subsection{Adding scalars to the dimensionally reduced system}\label{ssec:scalars}
Since we want to make contact with the Ginzburg-Landau phenomenological model for superconductivity,  we  introduce complex scalars which will play the role of order parameters. They shall have constant pairing amplitude $f^0$ but their phases, denoted by $\theta_{R,L}(x^i)$ to distinguish them from the auxiliary fields, are allowed to fluctuate:
\be
\Phi_{R,L}(x^i) = {f^0}_{R,L}\exp\left(i \theta_{R,L}(x^i)\right) \;.
\label{higgs}
\ee
 We shall take constant pairing amplitude $f_{R,L}^0$ (this being valid at very low temperatures $T$, $T \ll f_{R,L}^0$) and  fluctuations $\theta_{R,L}$ solely depending on $x_1$ and $x_2$.

We now promote the auxiliary field to dynamical ones. Note, from (\ref{eq:aux}), that
\begin{equation}\label{eq:aux1}
A_i(x)  \;=\; \partial_i[ - 2\pi \Theta(x^3) \, \xi_R({\mathbf x})  + 2 \pi \Theta(x^3-\ell) \, \xi_L({\mathbf x})] \;,
\end{equation}
where $\Theta$ is Heaviside's step function.
This expression suggests that, when promoting the auxiliary fields to dynamical ones, we let them have a non-trivial behaviour under the gauge transformations associated to $A_i$. Indeed, the shift of $\xi_L$ and $\xi_R$ by a function of the coordinates may be compensated by a  gauge transformation of $A_i$. Their effect on $A_i$  is compatible with the gauge transformation of fields which are the {\em phases\/} as the two complex scalar fields introduced in Eq.\eqref{higgs}.

 Using Eq.~\eqref{eq:aux1}, we see  that the original, $D=3$ Chern-Simons action, can be reduced to an effective $D=2$ effective action with an axionic coupling:
\be
S_{axion} = \frac1{4\pi}\int d^2x \frac{\theta_L(x^i) - \theta_R(x^i)}{2}\varepsilon^{ij} F_{ij}(x^i) \;.
\ee
This result is equivalent to the mapping between the topological superconductor in a \mbox{$D=3+1$} model and a \mbox{$D=4+1$} model which is
discussed in ref.~\cite{QiWi}. By an analogous analysis to the one presented in that reference, we shall now add to $S_{axion}$ the  covariant derivatives of scalars $\Phi_{R,L}$, writing an effective action $S_{eff}$, to make contact with a superconductor phenomenological free energy:
\be
S_{eff} = \frac1{4\pi}\int d^2x \frac{\theta_L - \theta_R}{2}\varepsilon^{ij} F_{ij}  +
\frac12{f_L^0}^2 (\partial_i\theta_L - 2A_i)^2 + \frac12{f_R^0}^2 (\partial_i\theta_R - 2A_i)^2 \;.
\label{super}
\ee
Note that the action $S_{eff}$ is similar to that introduced for the $D=4$ case in~\cite{QiWi}. The constants $f^0_{R,L}$ are dimensionless; from here on, we
take them to be equal to $1$.

 We see that, depending on the values of superconducting phase fluctuations
 $\theta_{R},\theta_L$ action $S_{eff} $ in~\eqref{super}, describes an ``ordinary'' superconductor ($\theta_{R}=\theta_L = 0$) or a ``topological''
 one ($\theta_{R}= \pi, \theta_L = 0$), the last one due to the first term containing the metric-independent $D=2$ integral. 

In order to clarify the result above, let first consider that there is just one fluctuating phase $\theta_L = \theta$ and write the field equation  associated
to the action \eqref{super}:
\be
 2 (\partial_i\theta_L - 2A_i) = \frac1{4\pi}\epsilon_{ij}\partial^j\theta - \frac1{e^2} \partial^j F_{ij} \;.
\label{zero}
\ee
Now, the left-hand side of this equation can be identified in $d=2$ dimensions with  the superconductivity current $j_\mu$ associated to the dynamical phase variable $\theta$ introduced through the complex scalar $\Phi$ in Eq.\eqref{higgs},
\be
{j_L}_i = 2   (\partial_i\theta_L - 2A_i) \;.
\label{tw}
\ee
Due to the axion coupling, we see that the current \eqref{tw}  leads  to a non-conserved $U(1)$ charge,
\be
\partial_i {j_L}_i  = \frac1{2\pi}\epsilon^{ij}\partial_i\partial_j\theta = \frac12\delta(x^1)\delta(x^2) \;.
\label{ano}
\ee
Now, the topological   density  associated to (zero radius) singular vortices\footnote{In Euclidean $d=2$ it would be more appropriate to call them a $d=2$ singular instantons}  takes the form
\be
\frac1{2\pi}\epsilon^{ij}\partial_i A_j = \delta^2(x) \;.
\label{medio}
\ee
Again, this result   can be associated to one half of such flux, this  indicating the presence of     Weyl-Majorana  fermions at the borders $x^3=L$ and $x_3 = R$ of the original  $d=3$ theory.

From Eq.\eqref{zero} (including the the contribution from the right-handed sector one has
\be
  (\partial_i\theta_L - 2A_i)  +  (\partial_i\theta_R - 2A_i) = 0
\ee
so that one can write
\be
A_i \approx\frac14 (  \theta_L +  \theta_R)
\ee
 Then we have for a loop encircling the $\theta_L$  vortex
\be
\oint_{\theta_L} A_i dx^i = \frac{\pi}{2}
\ee
and analogously, for the right-handed case
\be
\oint_{\theta_R} A_i dx^i = \frac{\pi}{2}
\ee
\subsection{Chiral vortices}
In the discussion above we have started with a theory in $D=3$ Euclidean dimensions and then reduced dimensions to $D=2$  dimensions that could be thought as spatial ones. Here we shall consider instead $D=2+1$ ($x^1, x^2, t$) Minkowski dimensions and then reduce to  $D=1+1$   ($x^1,t$) space time-dimensions.

In this case complex scalars still have  constant amplitudes but fluctuating   phases which may also depend on time: $\theta_{R,L} = \theta_{R,L}(x^1,t)$,
\be
\Phi_{R,L}(x^1,t) = f_{R,L}^0 \exp\left(i \theta_{R,L}(x^1,t)\right) \;.
\label{higgs1}
\ee
We now make the following gauge transformation
\be
A_a \to {\tilde A}_a + \partial_a \Lambda(x^2)
\ee
with $\Lambda$ chosen as
\be
\Lambda(x^2) = \frac1{2\ell} \left(\theta_L (\ell - x^2) - \theta_R x^2\right)
\ee
and
\bea
\theta_R(x^1,t) &\to& \theta_R(x^1,t)  + 2\Lambda(x^1,t, x^2=0)  \nonumber\\
\theta_L(x^1,t) &\to& \theta_L(x^1,t)  + 2\Lambda(x^1,t,x^2 =\ell)
\eea
With this, and repeating the analysis in the previous section  one has
\be
j_i = 2  (\epsilon_{ij}\partial_j  \theta - 2 A_i)
\ee
so that in this case
\be
\partial^i j_i = \frac12 \delta(x^1)\delta(t)
\ee
and hence
\be
\int\! dx^1\!dt \,\partial^i j_i = \frac12
\ee
so that this equation   coincides with  the  anomaly of Weyl fermions in the background of an electric field $F_{x^1\!t}$,
\be
\int\! dx^1\!dt \, \partial_i j^i = \frac1{4\pi} \int\! dx^1\!dt  F_{x^1t}
\ee
which coincides with which is half of the Chern number for the electric field

\section{Dimensional reduction from a fermionic theory}\label{sec:fer}
 In order to make contact with the results discussed above in the case of a fermionic theory in $2+1$ let us recall
that a Chern-Simons action is generated at the one-loop level effective action, as a consequence of the parity anomaly in $2+1$ dimensions, for a Dirac fermion coupled to a gauge field \cite{Redlich}.

The fact that Weyl fermions arise on the boundaries, can be justified in more that one concrete way, always within the context of a slab geometry.  As a first example, let us  then consider  a Dirac field which is confined to the same slab geometry discussed in the starting point of this work. Using an MIT bag-model~\cite{Johnson:1975zp} Euclidean action, concentrated on the slab, we have:
\begin{align}\label{eq:bag}
{\mathcal S}_f({\bar\psi},\psi;A) &=\; \int d^3x \,
{\mathcal L}_f ({\bar\psi},\psi;A) \;\;,\nonumber\\
{\mathcal L}_f({\bar\psi},\psi;A)  &=\; \Big[ {\bar\psi}(x)
 \frac{1}{2} \big( \overleftrightarrow{\not\!\partial} +
 i \not \!\! A + M \big) \psi \,+\, B \Big] \,
 \theta(x_3)\theta(\ell-x_3) \nonumber\\
 &+\;\frac{1}{2} {\bar\psi}(x) [\delta(x_3) + \delta(x_3 - \ell) ]\psi(x)
\end{align}
where $B$ is the bag constant (which plays not role here). The equations of motion that follow from this Lagrangian are:
\begin{align}\label{eq:bag1}
&\big[ \not\!\partial + i \not \!\!A(x) \,+\, M\big] \psi(x) \;=\;0
\;\;\; \forall x :\;\;\;  0 < x_3 < \ell \nonumber\\
&P_-\psi({\mathbf x},\ell) = 0 \;,\;\;
P_+ \psi({\mathbf x},0) = 0 \;, \;\;
P_\pm \equiv \frac{1 \pm \gamma_3}{2}
\end{align}
(and their Dirac adjoints).

We note that, in the fundamental representation of the Clifford algebra, the Dirac field above has two components,
\begin{equation}
\psi(x) \; \equiv \;
\left(\begin{array}{c}
\psi_+(x) \\
\psi_-(x)
\end{array} \right)
\;.
\end{equation}
Adopting the convention that $\gamma_i \equiv \sigma_i$ for $i =1,2,3$, with $\sigma_i$ denoting Pauli's matrices, the second line of (\ref{eq:bag1}) imply that at the $L$ and $R$ borders, the fields must have the form:
\begin{equation}
\psi(x_1,x_2,\ell) \; \equiv \;
\left(\begin{array}{c}
\psi_L(x_1,x_2) \\
0
\end{array} \right)
\;\;,\;\;\;
\psi(x_1,x_2,0) \; \equiv \;
\left(\begin{array}{c}
0 \\
\psi_R(x_1,x_2)
\end{array} \right)
\end{equation}
and
\begin{equation}
\bar{\psi}(x_1,x_2,\ell) \;=\;
\big( 0 \,,\, \bar{\psi}_L(x_1,x_2) \big)
\;,\;\;
\bar{\psi}(x_1,x_2,0) \;=\;
\big(\bar{\psi}_R(x_1,x_2) \,,\, 0 \big)
\;,
\end{equation}
for the Dirac adjoints.

Besides, at the borders, neither the mass term nor the term
involving $\gamma_3$ in the Dirac equation for the bag model,
act. Therefore, the one-component fields which appear at the boundary, satisfy the equations of motion
\begin{equation}
(D_1 + i D_2)_L \psi_L(x_1,x_2) \;=\;0 \;,\;\;
(D_1 - i D_2)_R \psi_R(x_1,x_2) \;=\;0
\end{equation}
with $(D_j)_L \equiv \partial_j + i A_j(x_1,x_2,\ell)$
and $(D_j)_R \equiv \partial_j + i A_j(x_1,x_2,0)$.

Recalling (\ref{eq:aux}), in the $A_3 \equiv 0$ gauge,
$A_i$ is a pure gauge, therefore it can be gauge away by a
standard (not chiral) gauge transformation of the fermions, except at the boundaries where we have Weyl fermions. Therefore, we pick up, on the borders, terms which are exactly like the ones we introduced by considering a Chern-Simons theory with borders.

One may wonder whether further terms for the scalar fields living on the boundary may be obtained also from a dimensional reduction of this fermionic system. Of course, parity conserving terms in the effective theory, like the kinetic terms for the auxiliary fields will require the introduction of both chiralities at the borders, like if one had massive modes.

Concerning fermion models in $D=2$ space-time dimensions, the path integral framework also allows to make contact at this point with recent results in \cite{K}. Indeed, let us consider the following Weyl fermion partition function $Z_{Weyl}$ in $D=1+1$:
\be
Z_{Weyl}[A,\theta] = \int D\bar\psi D\psi \exp \left(-\int d^2x L_{Weyl})\right)
\ee
where
\be
L_{Weyl} = \bar \psi ( i\sd\, - \sA +  M \exp\left(- 2\gamma_5 \theta(x_0,x_1)\right) \psi
\label{Li}
\ee
Here $A_\mu$ is a $U(1)$ external gauge field, $m$ is a constant with   dimensions and $\gamma_\mu$ are Euclidean Dirac matrices satisfying
\be
\{\gamma_\mu,\gamma_\nu\} = 2 \delta_{\mu\nu} \, , \hspace{1cm} \gamma_5 = i\gamma_0\gamma_1 \, , \hspace{1cm}\gamma_\mu\gamma_5 = i \varepsilon_{\mu\nu} \gamma_\nu \, , \hspace{1cm}  \epsilon_{01} = 1
\ee
We shall work with the following  representation for the $2\times 2$
Dirac matrices
\be
 \gamma_0 =  \left(
 \begin{array}{cc}
0 & 1 \\
1 & 0
\end{array}
\right)
\, , \;\;\;\;\;   \;\;\;\;\;    \gamma_1 =  \left(
 \begin{array}{cc}
0 & i \\
-i & 0
\end{array}
\right)
  \, ,  \;\;\;\;\;   \;\;\;\;\;  \gamma_5 =    \left(
\begin{array}{cc}
1 & 0 \\
0 & -1 \cr
\end{array}
\right)
\ee
Concerning the exponential term in the Dirac operator, it will play a role similar to the order parameter in Eq.\eqref{higgs} with the constant  $f_0$ identified with $m$ while $\theta$  is a $\vec x$-dependent fluctuation. Such a term corresponds to case B2 in the study  of   nonlinear sigma-models $\theta$-terms discussed in ref.\cite{Abanov}.

Let us  now perform a chiral change of fermionic variables in $Z_{Weyl}$
\be
\psi \to \exp \left( \gamma_5 \theta(x_0,x_1)\right)\psi \, ,\hspace{2 cm} \bar\psi \to  \bar \psi\exp \left(  \gamma_5 \theta(x_0,x_1)\right)
\label{q}
\ee
Of course this chiral transformation has an associated  Jacobian $J_5$ which can be reduced, working \`a la Fujikawa  \cite{Fujikawa} to the Jacobian relating the Grassmann coefficients in the expansion of the fermion variables. For the case of Weyl fermions one can follow the same steps as for the Dirac ones except that one has to include delta-functions among the Grassmann coefficients in order to define an appropriate Weyl fermion path integral measure. Such a procedure, described in detail in ref.\cite{GSS} leads to a
 Jacobian whose logarithm involves half of the Dirac operator  eigenvalues, with a double sign depending on the definition of Grassmann delta-functions
\be
\log J_{Weyl} = \pm \frac1{4\pi} \int d^2x A_\mu\varepsilon_{\mu\nu}\partial_\nu\theta \;.
\label{J5}
\ee
Note that the factor in front of the integral is one half the result for the Dirac fermions Jacobian associated to transformations \eqref{q}.

Using this result the  partition function $Z_{Weyl}$ takes the form
\be
Z_{Weyl} = \int D\bar\psi D\psi \exp \left(-\int d^2x L_{eff}\right)
\label {tive}
\ee
with $L_{eff}$ given by
\be
L_{eff} = \bar \psi ( i\sd + \sA  + M)\psi  \mp \frac1{4\pi} A_\mu\varepsilon_{\mu\nu}\partial_\nu\theta \;.
\label{ef}
\ee

Using eqs. \eqref{tive}-\eqref{ef} we can compute the fermion current v.e.v,
\be
  j_\mu(y)  \equiv  \frac1{Z_{Dirac}}\frac{\delta Z_{Dirac}}{\delta A^\mu(y)} = \frac1{4\pi}\varepsilon_{\mu\nu}\partial_\nu\theta(y)
\ee
so that we obtain  for the Weyl fermion current
\be
\partial_\mu   j_\mu(y) = \frac 12\delta^2(y)\;.
\ee

Finally, we note that a Dirac field with a different kind of slab geometry
may be considered, which also yields a similar effective theory. Indeed, we can introduce a Dirac action in $2+1$ dimensions with a space-dependent mass:
\begin{equation}
{\mathcal S}_f(\bar{\psi},\psi;A)\;=\;\int d^3x \, \bar{\psi} \big[\not\!\!D + M(x_2) \big]\psi \;.
\end{equation}
Using a mass profile which changes sign twice, precisely at the locii of the regions where one wants to localize the fermions,
\begin{equation}
M(x_2)\;=\; \left\{
\begin{array}{rcl}
M & {\rm if} &  x_2 < 0 \; {\rm or} \; x_2 > \ell \\
- M & {\rm if} &  0 < x_2 < \ell \;,
\end{array}
\right.
\end{equation}
where $M$ is a constant.
The fact that there are $D=2$ fermions localized at $x_2 =0$ and $x_2=\ell$ may be seen from an application of Callan-Harvey mechanism \cite{CallanHarvey}, noting that the jump in the mass has opposite signs at those points; therefore, there will be fermions of opposite chiralities: a $L$ at $x_2=\ell$ and right at $x_2 =0$.

A way to make that more transparent is to expand the fermion field in terms of the eigenstates of a suitable self-adjoint operator. A convenient one is the combination ${\mathcal H} \equiv {\mathcal D}^\dagger {\mathcal D}$, where
\begin{equation}
{\mathcal D} \;\equiv\; \not\!\!D + M(x_2)  \;.
\end{equation}
Assuming that $a_2$ is independent of $x_0$ and $x_1$, and that $A_1$ and $A_2$ are independent of $x_2$, we see that:
\begin{equation}
{\mathcal H} \;=\; ( a^\dagger a - \not\!\!D_\shortparallel^2) P_+\,+\,
( a a^\dagger - \not\!\!D_\shortparallel^2) P_- \;,
\end{equation}
where
\begin{equation}
a \,=\, D_2 \,+\, M(x_2) \;,\;\;
a^\dagger \,=\, - D_2 \,+\, M(x_2) \;\;\;,\;\;\;\;
\not\!\!D_\shortparallel \,\equiv\, \gamma_0 D_0 + \gamma_1 D_1 \;.
\end{equation}
The Weyl fermions corresponds to the zero modes of the $a$ and $a^\dagger$ operators, which for the considered mass profile sit at $x_2=\ell$ and $x_2=0$, respectively.


\section{Summary and discussion}\label{sec:conc}
In conclusion, in this work we have discussed models in $d=1+1$ and $d=2+1$ dimensions  that presently show growing interest in the study of topological order in quantum field theory and condensed matter physics.

 The path-integral approach that we have followed is a useful tool to understand the symmetry behavior of quantum field theories and the possible existence of anomalies whenever the path-integral measure is not invariant under the classical symmetry; in that case the resulting Jacobian discloses the topological {character  of a model}.

In section 2 we start by studying  an Abelian Chern-Simons theory in $2+1$ dimensions defined in a manifold which has a non-trivial boundary:  two parallel planes which were introduced in the partition function using two scalar  Lagrange multipliers. As a result  we were able to relate the CS current  with the Weyl fermion anomaly, this being  interpreted as  the existence of two Majorana-Weyl fermions with opposite helicities at the edges as discussed in \cite{QiWi} for the case of a $d= 4+1$ CS term.

In section 3 we have promoted the previous section auxiliary fields   to dynamical phases  $\theta_{LR}$  with  an appropriate
  action \eqref{super}   Then we have shown that the superconductivity current associated to the   $\theta_{LR}$ corresponds to the presence of Weyl-Majorna fermions at the borders.
We have also considered the model in Minkowski space and then reduce the action to the $(x^1,t)$ space-time.

Finally in section 4 we considered a Dirac field in $d=2+1$ space-time which is confined to the same slab geometry discussed above with action \eqref{eq:bag}. As a result we found the same results which agree with those obtained previously for the Chern-Simons model with borders. We also discussed a  $d= 1+1$ Weyl fermion model but in this case instead of the bag term above, we considered a ``mass'' term
$M \exp\left(- 2\gamma_5) \theta(x_0,x_1)\right)$, already studied for the case of fermionic $\sigma$-models in \cite{Abanov}. An appropriate chiral change of variables eliminates the $\theta$ dependence in the action but the associated Jacobian introduces an axion coupling as the one discussed above.

A different kind of slab geometry is discussed as a final example where fermions in $d=2+1$ dimensions  have a space dependent mass such that, using the Callan-Harvey mechanism, one ends with $1+1$ dimensional  Weyl fermions sitting at the ends of the regions where   one want to localize the fermions.

Let us end by noting that working at finite temperature could lead to interesting effect in the results discussed above. In the case of fermions in $2+1$ dimensions, after some controversy about gauge invariance of the Chern-Simons effective action arising from fermion integration it was shown   that  the correct calculation at $T \ne 0$  temperature leads to a non-extensive effective action in Euclidean time but
extensive quantity in Euclidean space \cite{FRS1}.
Another interesting issue concerns the so-called Witten effect \cite{Witten} in which the presence of a CP violating $\theta$ term make  dyons acquire a $\theta$ dependent electric charge. Although Witten formula does not change at $T \ne 0$ \cite{LGS} , it does change when fermions are coupled to the theory \cite{GP}. We expect to study more thoroughly these issues in forthcoming work.

\vspace{1.2 cm}

\noindent{\bf{Acknowledgments:}}
F.A.S. is financially supported by PIP-CONICET (grant PIP688) and UNLP  grants.
C.D.F. acknowledges support by CONICET, ANPCyT and UNCuyo.

\end{document}